\documentclass[12pt,a4paper]{article}
\usepackage{epsfig}
\usepackage{amstex}
\usepackage{amssymb}

\newcommand{\erfc}{{\mbox{erfc}}}

\newcommand{\IK}{{\mathbb K}}

\newcommand{\ZZ}{{\mathbb{Z}}}

\newcommand{\VECk}{{\mathbf{k}}}

\newcommand{\VECm}{{\mathbf{m}}}
\newcommand{\VECn}{{\mathbf{n}}}
\newcommand{\VECr}{{\mathbf{r}}}

\begin{document}
\vspace*{-3cm}
\hfill Report M-98-146\\[2cm]

\centerline{{\LARGE \bf Polyelectrolytes in Solution-}}
\vspace{0.3cm}
\centerline{\LARGE \bf Recent Computer Simulations}
\vspace{0.5cm}
\centerline{\large {\em Christian Holm\/} and
                   {\em Kurt Kremer\/}} 
\vspace{0.2cm}
\centerline{{\small Max-Planck-Institut f\"ur Polymerforschung
                      }}
\vspace{-0.1cm}
\centerline{{\small Ackermannweg 10, 55128 Mainz, Germany }}
\vspace{0.5cm}
\begin{abstract}
We present a short overview over recent MD simulations of systems of fully
flexible polyelectrolyte chains with explicitly treated counter ions using the
full 
Coulomb potential. The main emphasis is given on the conformational properties
of the polymers, with a short discussion on counter ion condensation.
\end{abstract}
               \section{INTRODUCTION}
Polyelectrolytes represent a broad and very interesting class of
materials \cite{pe_review} that enjoys an increasing attention in the
scientific community.
Even though the theory of neutral polymer systems is well developed,
polyelectrolytes remain one of the least understood states of the soft matter area.
Theoretically this is mainly due to the long range Coulomb interaction.
Simple scaling theories, which have been proven so successfully in neutral
polymer theory, have to deal with the additional length scales set by the
electrostatic interaction. It is the delicate interplay between electrostatic
interaction, and the conformational degrees of freedom, which in turn are
governed by a host of short range interactions, which proves to be a difficult
question for the theory.
This results in essence in the fact that the analytical theory
is only able to treat two limiting cases, namely the case of high salt excess,
resulting in effectively screening down the electrostatic interaction to treat
it as a perturbation, or the case of an overwhelming dominance of the Coulomb
force, which results in a strongly elongated chain.
Unfortunately it is just the intermediate case, which proves to be the most
interesting regime in terms of application, experiment and theory.

How difficult it is for the analytical theory to treat
this intermediate regime could be
seen in the recent controversy over the correct exponent $y$, which governs
the variation of the electrostatic persistence length $L_e$ with Debye
screening length $r_D$. The theory is already simplified because 
instead of the full Coulomb interaction it assumes a screened Debye-H\"uckel
potential 
$U_{DH} (r) = \lambda _B k_B T \frac{\exp (r/r_D)}{r}$, 
where $\lambda _B$ is the Bjerrum length given by
$\lambda_B = \frac{e^2}{4\pi\epsilon_s \epsilon_r k_B T}$, with $e$ denoting
the elementary charge and $\epsilon_s, \epsilon_r$ are the dielectric
constants. The two competing theories both predicts $L_E \propto r_D ^y$, 
however the exponents are $y=1$ or $y=2$,
depending which perturbation approach one believes more. To test the theoretical
approaches, a series of molecular dynamics (MD) and Monte Carlo (MC)
simulations have been presented \cite{um} with the result, that neither
theoretical approach is correct in the intermediate regime, but that the
exponent $y$ can vary continuously between 0.5 and 2. 
Only very recently a field theoretic renormalization group analysis was able
to shed some additional light onto this problem\cite{tanni}.

A serious drawback of the Debye-H\"uckel theory is that it 
presupposes only weak variations in the density of the counterions, and
neglects all correlation effects. One can therefor doubt if this potential is
able to describe the chain conformations adequately \cite{mark}. 
Also counterion condensation can not properly be handled.
To tackle these questions computer simulations provide the necessary tools,
because here one does not need to rely on excessive simplifications.
One can treat the counterions explicitly and can also utilize the full
long range Coulomb potential. We try to summarize here recent 
results of MD simulations on {\it systems} of polyelectrolyte
chains in solution. Details of the studies can be found in the original
literature \cite{kk,phd_micka97,poster_DH97}
               \section{MODEL}
Our model of a flexible polyelectrolyte chain consists of $N_p$
bead-chain polymers with $N_m$ monomers which are located in a simulation box
with periodic boundary conditions (3D torus). From
these monomers a fraction $f$ is 
monovalently charged. The number of counterions, $N_c$, with valence $v$ is
then chosen such that the overall system is electrically neutral. If we have,
for example, $N_q$ charges on a polymer chain, then the number of counterions
with valence $v$ is given by $N_c = \frac{N_p N_q}{v}$, and the total number
of charges in the system is $N_{tq}= 2N_c$. All {\it hydrophilic}
monomers are given an effective size through a pure repulsive Lennard-Jones
potential, representing thus a polymer chain in a good solvent:
\begin{equation}
U_{LJ}^{r}(r) = \left\{ \begin{array}{c@{\quad : \quad}l}
4\varepsilon _{LJ} \left[ \left(\frac{\sigma}{r}\right)^{12} - 
\left(\frac{\sigma}{r}\right)^{6} + \frac{1}{4} \right] & 
r \le r_{min}= 2^{1/6} \sigma\\
0 & \mbox{otherwise}
\end{array} \right.
\label{eq:ljhp}
\end{equation}
All chain monomers, which model the {\it hydrophobic} or 
poor solvent case, interact via a standard Lennard-Jones 
potential with attractive part:
\begin{equation}
U_{LJ}^{a}(r) = \left\{ \begin{array}{c@{\quad : \quad}l}
4\varepsilon _{LJ} \left[ \left(\frac{\sigma}{r}\right)^{12} - 
\left(\frac{\sigma}{r}\right)^{6} \right] & r \le r_{c}= 2.5 \sigma \\
0 & r \ge r_{c}= 2.5 \sigma 
\end{array} \right.
\label{eq:ljhb}
\end{equation}
All {\it chain} monomers are in addition connected by the FENE
(finitely extended nonlinear elastic) bond potential,
\begin{equation}
U_{\rm FENE}(r) = -\frac{1}{2} k R_0^2 \ln \left(1 - \frac{r^2}{R_0^2}\right).
\end{equation}
All {\it charged} monomers posses in addition the full
Coulomb energy
\begin{equation}
E_c (r_{ij}) = \frac{ \lambda_B k_B T v_i v_j}{r_{ij}},
\end{equation}
where $v_i$ is the valence of the $i^{th}$ charged monomer.

The electrostatic energy of the box is calculated with the Ewald formula
\begin{equation}\label{EwaldAnteile}
E_c = E^{(r)} + E^{(k)} + E^{(s)}
\end{equation}
where the contribution from real space $E^{(r)}$, the
contribution from reciprocal space $E^{(k)}$, and the self 
energy $E^{(s)}$ are given by
\begin{eqnarray}
E^{(r)} & = & \frac{1}{2} \sum_{i,j} \sum_{\VECm\in\ZZ^{3}}^{\prime}
q_{i}q_{j} \frac{\erfc(\alpha|\VECr_{ij}+\VECm L|)}
{|\VECr_{ij}+\VECm L|} \label{Realraumanteil} \\
E^{(k)} & = & \frac{1}{2}\frac{1}{L^{3}} \sum_{\VECk\ne 0}
\frac{4\pi}{k^{2}} e^{-k^{2}/4\alpha^{2}}
|\tilde{\rho}(\VECk)|^{2} \label{Impulsraumanteil} \\
E^{(s)} & = & -\frac{\alpha}{\sqrt{\pi}}\sum_{i}q_{i}^{2}
\label{Selbstenergie}
\end{eqnarray}
and the Fourier transformed charge density $\tilde{\rho}(\VECk)$
is defined as
$\tilde{\rho}(\VECk) = 
\sum_{j=1}^{N_{q}}q_{j}\;e^{-i\,\VECk\cdot\VECr_{j}}$.

The Ewald parameter $\alpha$, which has the dimension of an inverse length, 
tunes the relative weight of the real 
space and the reciprocal space contribution, but the final result 
is of course independent of $\alpha$. The $\VECk$-vectors form 
the discrete set $\{2\pi\VECn/L:\VECn\in\ZZ^{3}\}$, and we use tin-foil
boundary conditions, hence there is no dipole correction term. The
standard Ewald method has computational effort at best of 
${\cal O}(N_{tq}^{3/2})$,  
which limits the size of the samples. To overcome this limitation, we use 
for $E^{(k)}$ a
particle mesh Ewald algorithm (PME) that uses discrete charge assignments and 
fast
Fourier transforms (FFT), and accordingly scales for $E^{(k)}$ as 
${\cal O}(N_{tq} (\log N_{tq})^{3/2})$ \cite{pme}. Note, however, that 
this is not the most effective algorithm, as was shown recently \cite{dh98a}.

In all our performed molecular dynamics (MD) simulations we had no salt ions 
and no explicit solvent molecules.
The solvent,however, was implicitly taken into account via a Langevin 
thermostat with
damping constant $\Gamma = \tau^{-1}$, with time step $0.00125\tau$ at constant
temperature $k_B T=1\epsilon$. The number of MD steps was chosen such 
that the typical observables like the end-to-end distance 
$R_{E}= \sqrt{\langle \vec R_{E}^{2} \rangle}$ or the radius of 
gyration $R_{G}$ had sufficiently relaxed, which happened usually 
after 500\,000 up to 5\,000\,000 MD steps.
               \section{STRONGLY CHARGED POLYELECTROLYTES IN GOOD SOLVENT}
         \subsection{MONOVALENT COUNTERIONS}
The first investigation of totally flexible many chain 
polyelectrolyte systems in good solvent with 
explicit monovalent counterions was performed already some years ago
\cite{kk}. The simulations were done mostly with
systems of 8 or 16 chains with $N_m$ = 16, 32, and 64. Instead of the PME
algorithm a spherical approximation in a truncated octahedral simulation box
was used, which is for smaller values than $N_{tq} \approx 500$ faster than the
PME method. More details of the whole study can be found in \cite{kk}.

In this work it was demonstrated that the simple bead-spring model can
actually be 
compared to real polyelectrolytes, because the experimental values for the
osmotic pressure and the maximum position in the interchain structure factor
are successfully being reproduced.
One of the important findings was that the rodlike chain is found to be a
rarity already in the dilute limit. Counterion condensation can dramatically
shrink the polyelectrolyte chain. The end-to-end distance shortens
significantly as the density increases from the dilute saturation value to the
overlap value. The chain structure is highly asymmetric at the very dilute
saturation density and the scaling with respect to $N_m$ is asymmetric, but as
the overlap density is approached, the structure is less asymmetric and the
scaling becomes approximately symmetric. On long length scales the chain
structure continuously changes from very elongated to neutrallike coils. Yet,
on short length scales, the chain structure is density independent and
elongated more than neutral chains.
\begin{figure}[tbp]
  \begin{center}
    \leavevmode
    \epsfig{file=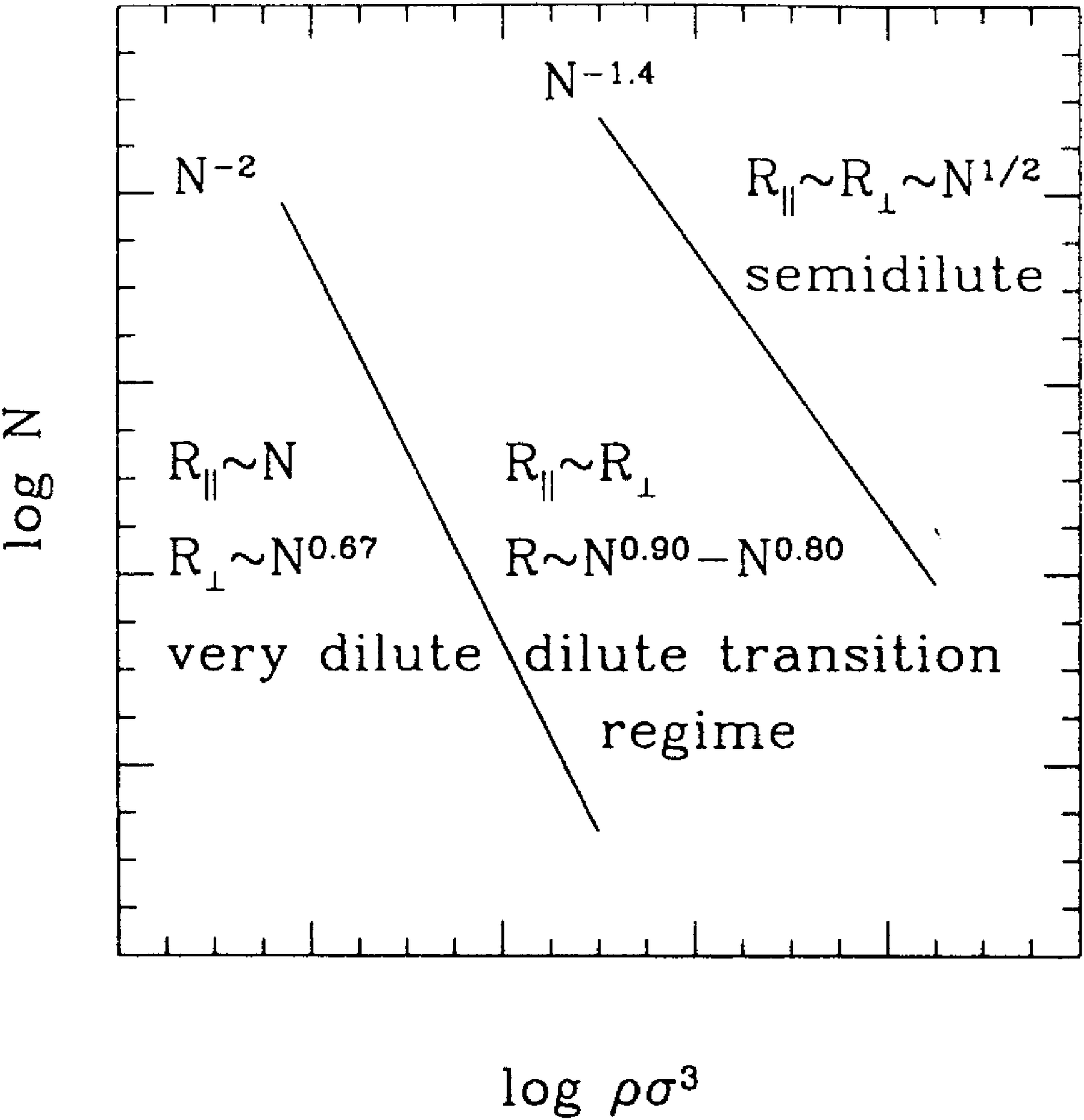,width=7.7 cm}
    \caption{A structure diagram for $\lambda_B = 0.833\sigma$, which shows the
    dilute limit, the contraction below the overlap density $\rho ^*$, and the
    completely screened limit at large $\rho$.}
    \label{kk_fig_1}
  \end{center}
\end{figure}
The findings for the single Bjerrum length $\lambda_B = 0.883 \sigma$ are
summarized in Fig.~\ref{kk_fig_1}. It was found that in the dilute limit 
the scaling for the extension perpendicular to the chain was $R_\perp \propto
N^{0.65-0.70}$, and 
for the extension parallel it was $R_\parallel \propto N^{0.90-1.00}$. Near the
density, where the rodlike chains in disordered solution would overlap, $\rho
\sim N^{-2}$, $R_\perp$ starts to grow on the expense of $R_\parallel$ until
at the overlap density $\rho ^*$ the effective 
exponent is about 0.82. This transition
regime ranges from $\rho \sim N^{-2}$ to about $\rho \sim N^{-1.4}$ where the
coils start to overlap and one eventually reaches $\nu = 1/2$ in the
semidilute regime. The exponents reported should not necessarily be taken as
asymptotic ($N \rightarrow \infty$), however they should be relevant for many
experimentally systems.

In a recent study \cite{phd_micka97} a large MD simulation of 200 chains with
$N_m = 32$ was 
performed to determine if with these large systems ($N_{tq} = 12\,800$)
any large scale ordering phenomena could be observed. The idea of the onset of
ordering is one of the possible explanations for the Fuoss-Strauss
effect in the reduced viscosity $\eta_r$, which
shows a dramatic increase and subsequently a decrease of $\eta_r$, as the
polymer concentration is decreased.
The pure MD simulation showed so far no sign of correlations or ordering,
however it was found that the center of mass diffusion of the monomers was
just too slow to move the monomers far enough. It needs definitely accelerated
algorithms (MC or MD) to settle this question in 
a large scale computer simulation, which will be pursued in the future.
         \subsection{MULTIVALENT COUNTERIONS}
\begin{figure}[bt]
  \begin{center}
    \leavevmode
    \epsfig{file=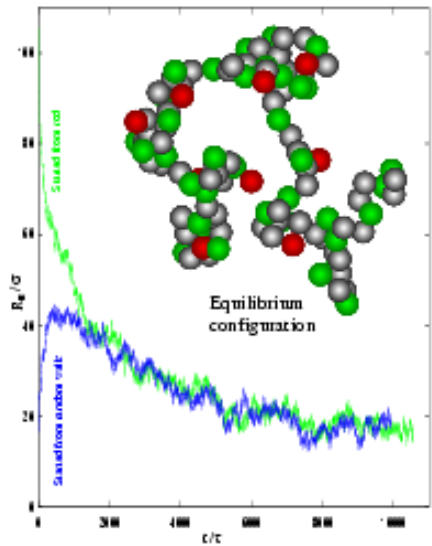,width=7.7 cm}
   \caption{Time series of the end-to-end distance $R_E$ versus
    simulation 
    time $t/\tau$, plus a typical equilibrium configuration for a trivalent
    system at $\rho  = 10^{-5}\sigma^{-3}$.} 
    \label{holm_fig_timeseries}
  \end{center}
\end{figure}

\begin{figure}[bt]
  \begin{center}
    \leavevmode
    \epsfig{file=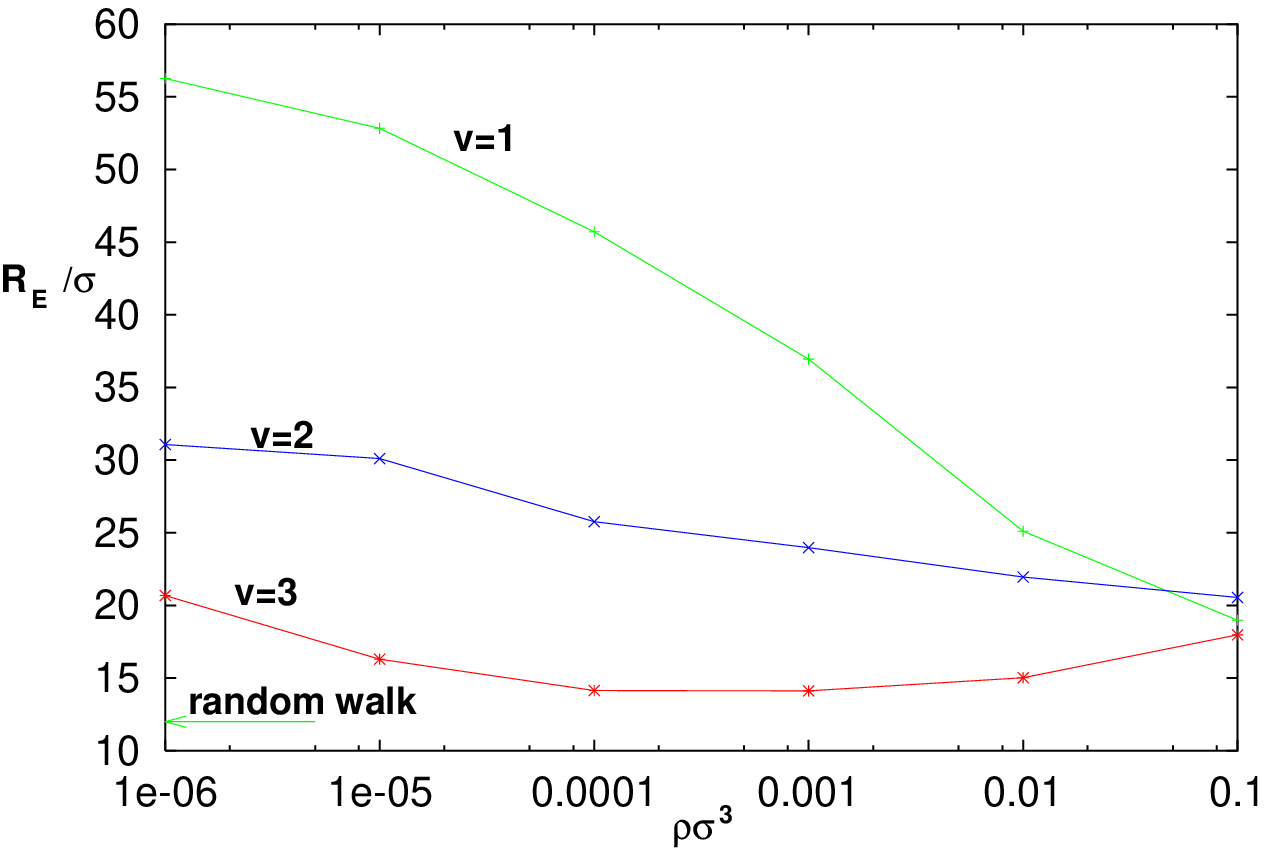,width=7.7 cm}
    \epsfig{file=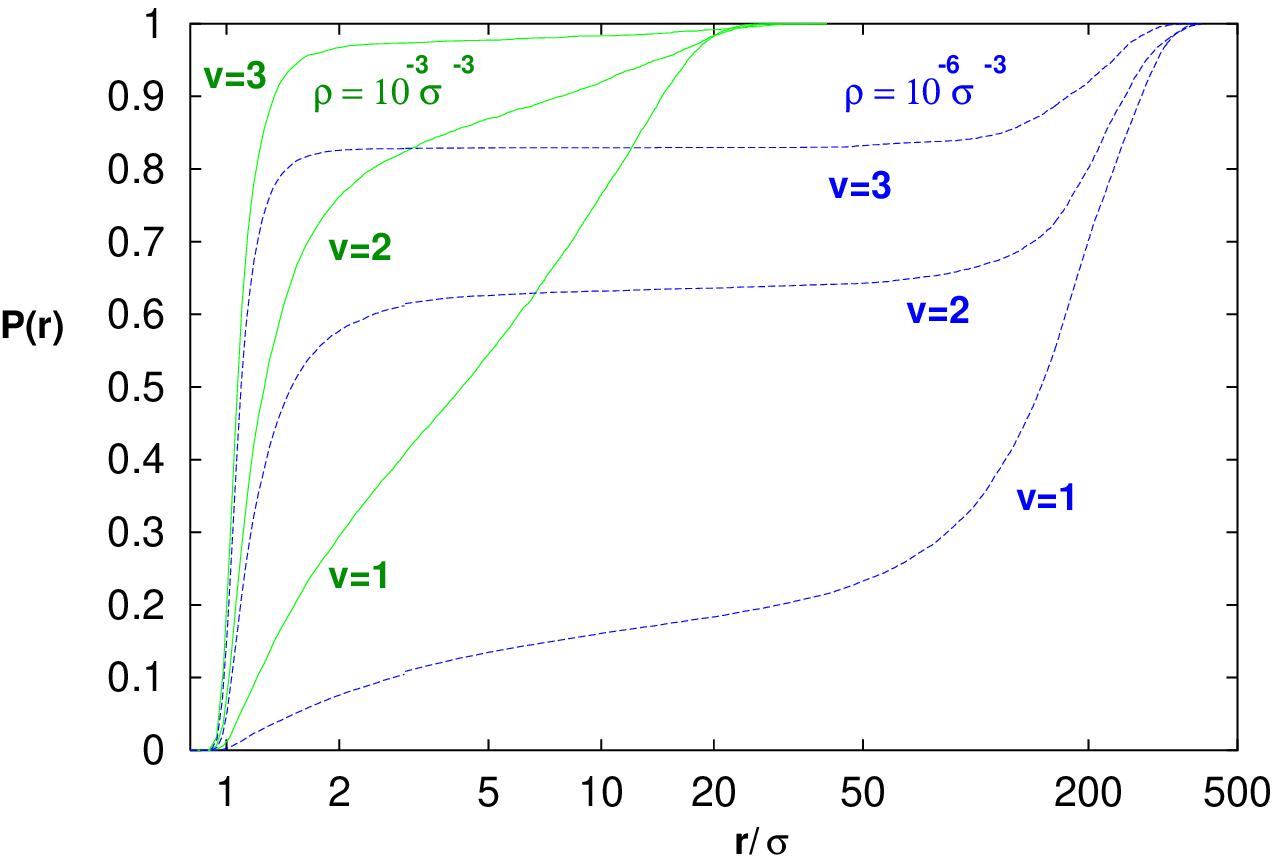,width=7.7 cm}
    \caption{Left: $R_E$ vs. $\rho$ for valences $v = 1...3$; right:
    integrated ion distribution 
    function P(r) vs. distance $r$ for $\rho = 10^{-3} \sigma^{-3}$ and $\rho
    = 10^{-6} \sigma^{-3}$, each for valences $v = 1...3$.} 
    \label{holm_fig_multi}
  \end{center}
\end{figure}

In this set of first simulations we usually considered eight chains in the 
central box with a chain length of
$N_{m}=106$ and a charge fraction of $f=1/3$. 
We varied the valence of the counterions from one to three,
and chose always $\lambda_{B}= 3 \sigma$, thus the monovalent 
systems were just below the critical Manning threshold. The weighted 
density of charged particles, 
$\rho _{c} = \sum _{i} ^{N_{tq}} \frac{v_{i}}{L^{3}}$
was varied from 
$10^{-1}\sigma^{-3} \dots 10^{-6}\sigma^{-3}$.
The starting configurations were set up as rod configurations with randomly
distributed counterions, or as a random walk, if the rods would not fit in the
central box. That both initial configurations 
lead to the same equilibrium value for $R_E$ can
be seen in Fig.~\ref{holm_fig_timeseries}.
Another point to note is, that both time series converge very rapidly, even
before they are fully equilibrated. This is due to the fact, that the
Coulombic repulsion of the chain monomers provide a very fast initial
rearrangement of the chain, but that the
equilibrium conformation is only reached, after the counterion cloud has
itself arranged around the polymers.
If one looks on the end-to-end distance $R_E$ as a function of density for the
different valences, see left side of Fig.~\ref{holm_fig_multi}, one
notices that for high polymer concentrations all 
three valences have basically the same $R_E$, they show SAW behavior. For the
semi-dilute range the monovalent systems elongate very fast, whereas the
divalent systems elongate much slower, and the trivalent system actually 
{\it contracts}. This is due to the fact that a trivalent counterion can bind
more than one charged chain monomer, so that the chain can ``wrap around'' the
counterion, or the counterion can even bridge two far apart regions of
the chain, as can be seen in the configuration in the upper part of 
Fig.~\ref{holm_fig_timeseries}.
For lower density the trivalent chains elongate much slower, so
that they will reach a rod like configuration at much lower
concentrations. Even at this dilution the chains still show much structure due
to the discrete counterions.

Another interesting quantity to look at is the fraction of counterions found
up to a distance r, which we denote by the {\it integrated}
ion distribution function $P(r)$,
shown on the right side of Fig.~\ref{holm_fig_multi}. Especially for
low concentrations, this function gives a good estimate to determine the
fraction of {\it condensed} counterions, because one can clearly see a plateau
value for $P(r)$, meaning that the ions can be separated in bulk counterions
far apart from the polymer, and a distinct layer of condensed counterions.
For higher concentrations and lower valences, however, this distinctions gets
less pronounced. One observes that the condensed fraction of counterions,
which we denote by $C$,
depends strongly on the density and valence. With increasing density and
valence the condensed fraction increases, and assumes for the highest
investigated density $\rho = 10^{-1}\sigma^{-3}$ the value $C=1$ for all three
valences. The Manning predictions for these systems are $C=0 (v=1)$, 
because for the monovalent systems we are slightly below the critical Manning
parameter, and $C\approx 1/2 (v=2)$ and $C \approx 1/3 (v=3)$ for the other
valences. 

A third interesting aspect to look at is the counterion exchange dynamics. If
one defines the set of condensed ions as $\IK_{d}^j(t)$, which contains all
ions being within distance $d$ on polymer $j$ at time $t$, 
then one can define a conditional condensation fraction $c_d((t|t_0)$ as
follows:
\begin{equation}
  \label{eq:con_condens_fraction}
    c_{d}\left(t|t_{0}\right) := 
  \left\langle\frac{|\IK_{d}^{j}(t) \cap \IK_{d}^{j}(t_{0})|}
    {|\IK_{d}^{j}(t_{0})|}\right\rangle_{j}
\end{equation}
This equation simply means, count the number of ions which are on time
$t$ as well as on time $t_0$ condensed on polymer $j$, normalize by the number
of condensed ions on polymer $j$ at time $t_0$, and average then over all
polymer chains $j$. The normalization
is such, that $c_{d}(t_{0}|t_{0})=1$. Fitting this quantity to an exponential
Ansatz $c_d(t|t_0) \sim A + B\exp{-t/t_{\rm relax}}$ gives information about
the exchange dynamics of the counterions. We find that with increasing
valence at constant density the mobility decreases, and that the mobility
also decreases with decreasing density at constant valence. The value of
$t_{\rm relax}$ is of the order of $10\,000 \tau$, showing the very slow
dynamics at low densities.

               \section{POLYELECTROLYTES IN POOR SOLVENTS}
The investigations so far dealt only with polyelectrolytes in good solvent, to
facilitate the comparison to known systems such as neutral polymer
systems. Most experimentally investigated polyelectrolytes, however, 
posses a
hydrophobic backbone. This results in a competition between the 
solvent quality, the Coulombic repulsion, and the entropic degrees of freedom,
which can lead to totally different behavior\cite{phd_micka97}. The
simulations presented below used 16 chains of length $N_m =94$, with a 
charge fraction of
$f=1/3$, and monovalent counterions.
The hydrophobic interaction strength was tuned by means of the 
Lennard-Jones parameter
$\varepsilon _{LJ}$ of Eq. (\ref{eq:ljhb}) that was chosen such that the 
finite chain shows an effective random
walk behavior ($\varepsilon_{LJ}=1.5$) for the density 
$\rho = 10^{-4}\sigma^{-3}$.

\begin{figure}[htbp]
  \begin{center}
    \leavevmode
    \epsfig{file=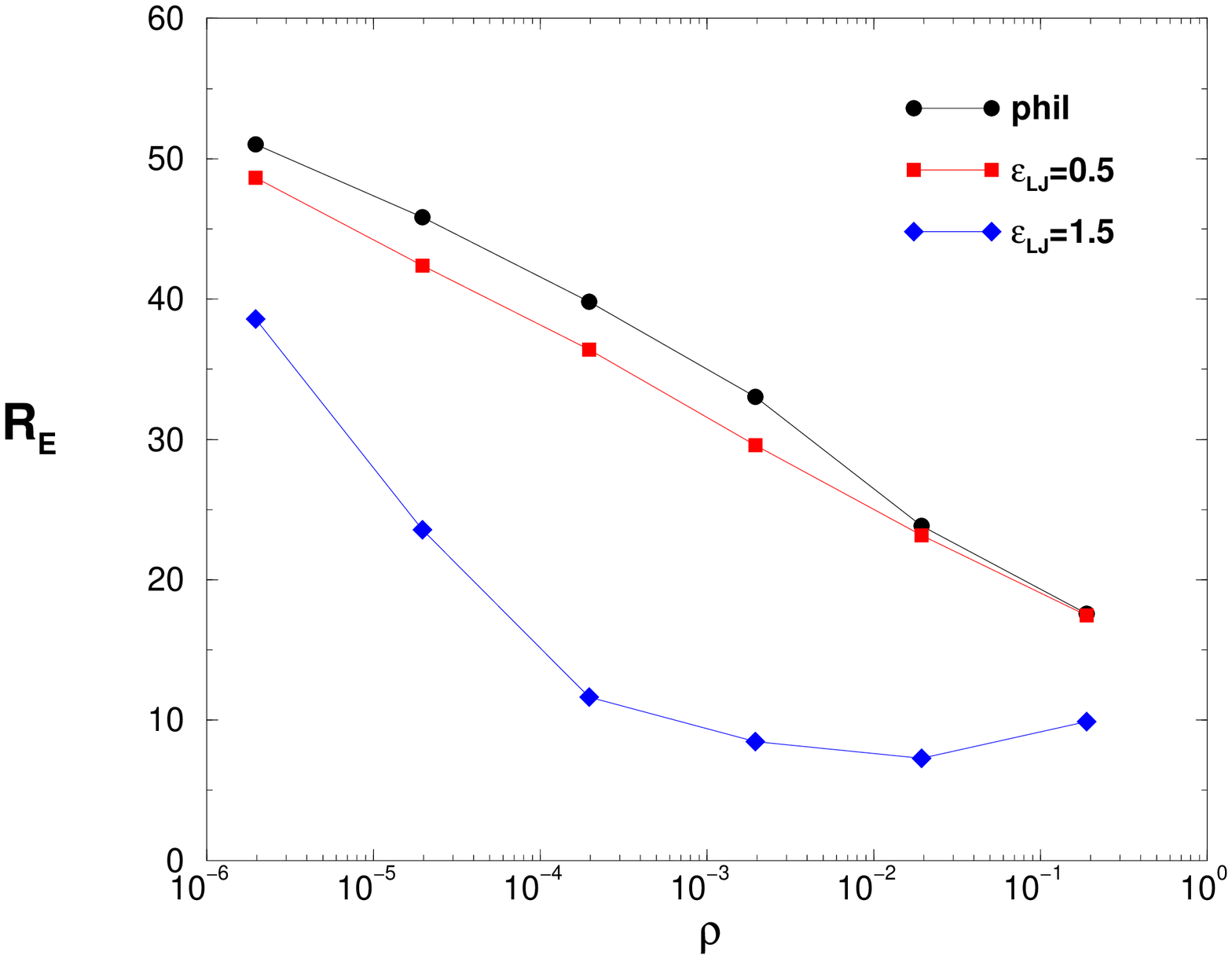,width=7.5cm} 
    \epsfig{file=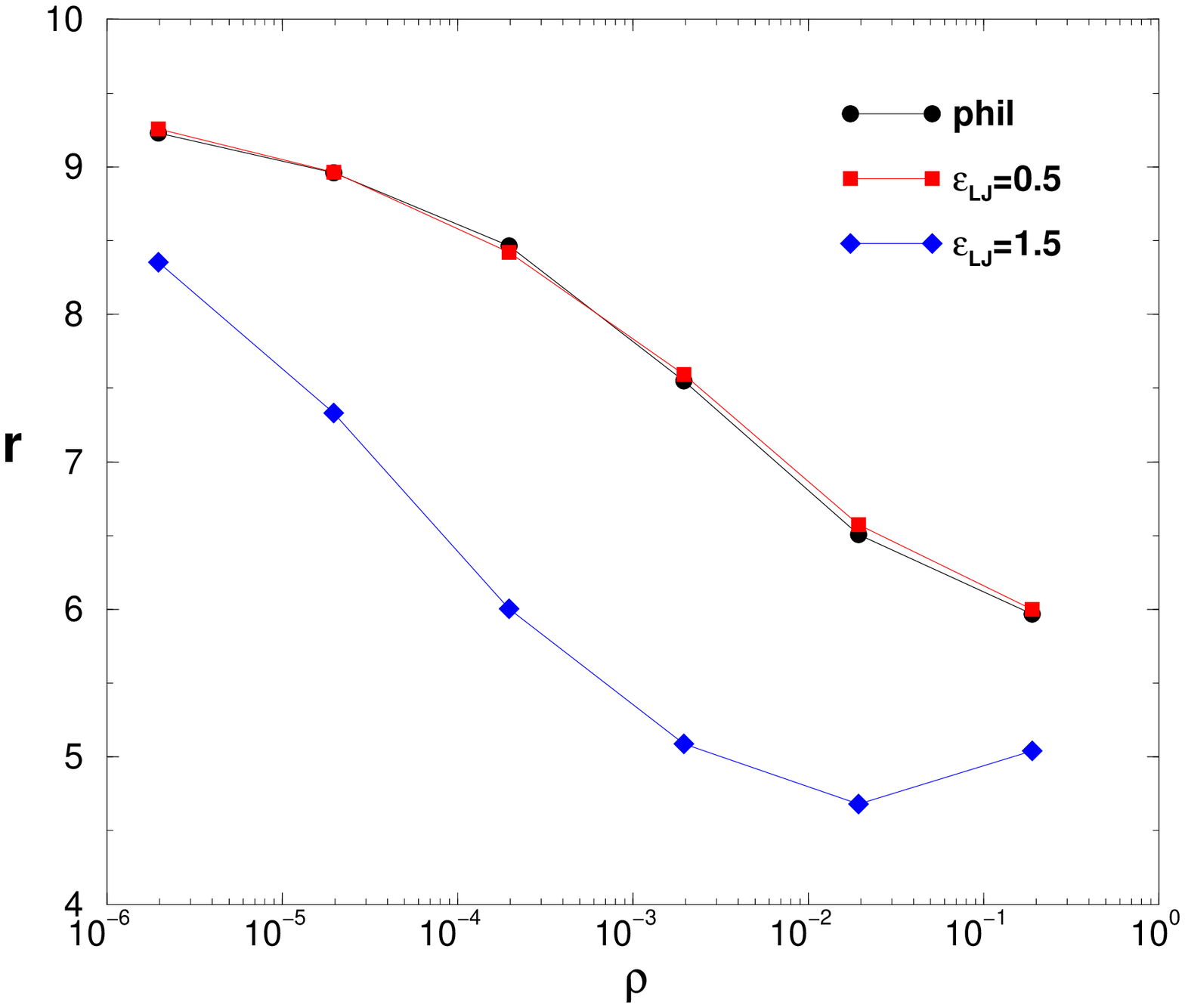,width=7.5cm}
    \caption{ ${\rm R_E}$ (left) and r (right) versus density $\rho$ for
    hydrophilic 
    (phil), weak hydrophobic ($\epsilon_{LJ} = 0.5$), and strongly hydrophobic
    ($\epsilon_{LJ} =1.5$) chains} 
    \label{micka_fig_re}
  \end{center}
\end{figure}

\begin{figure}[tbp]
  \begin{center}
    \leavevmode
     \epsfig{file=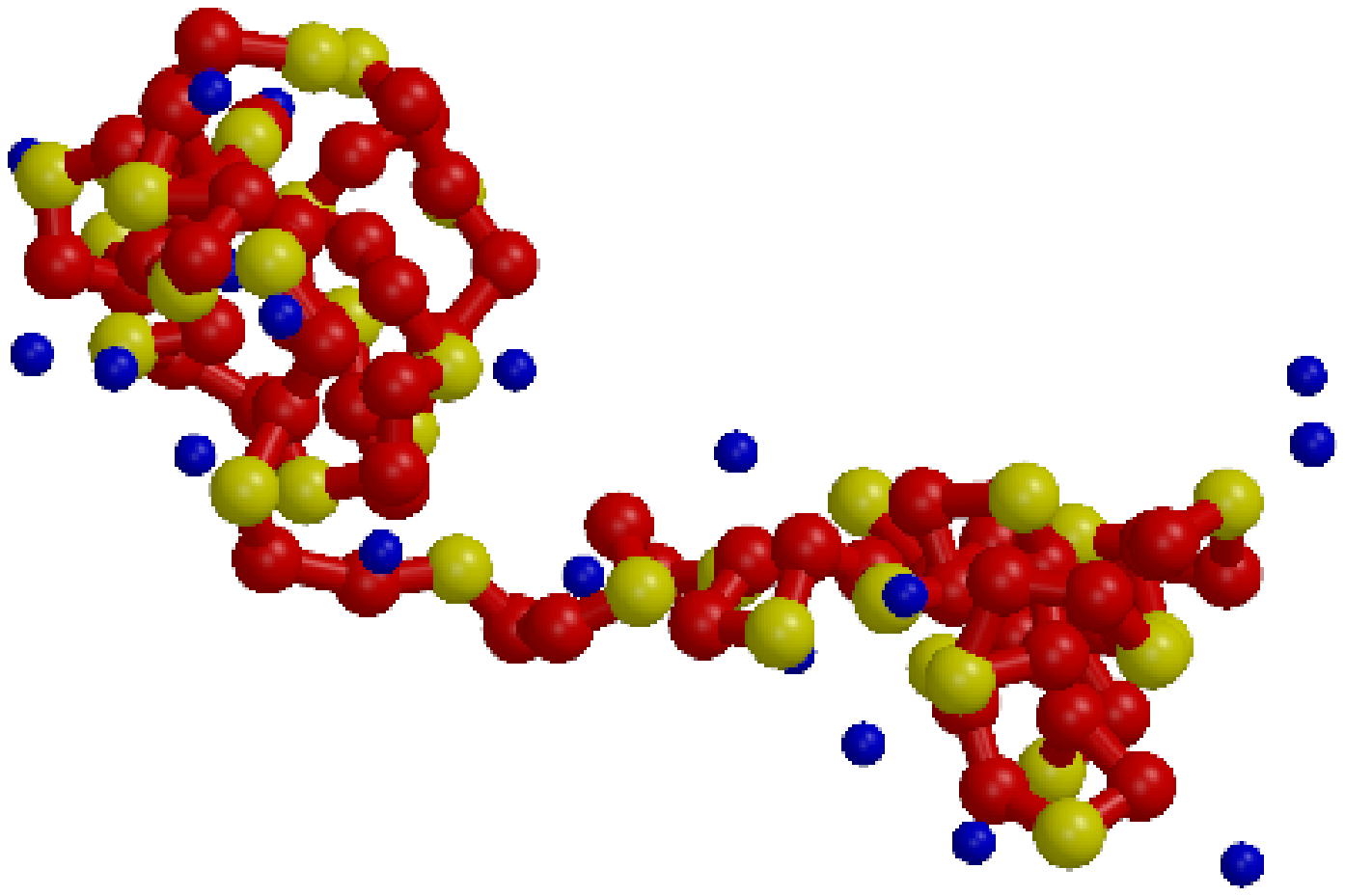,width=7.5 cm}
     \epsfig{file=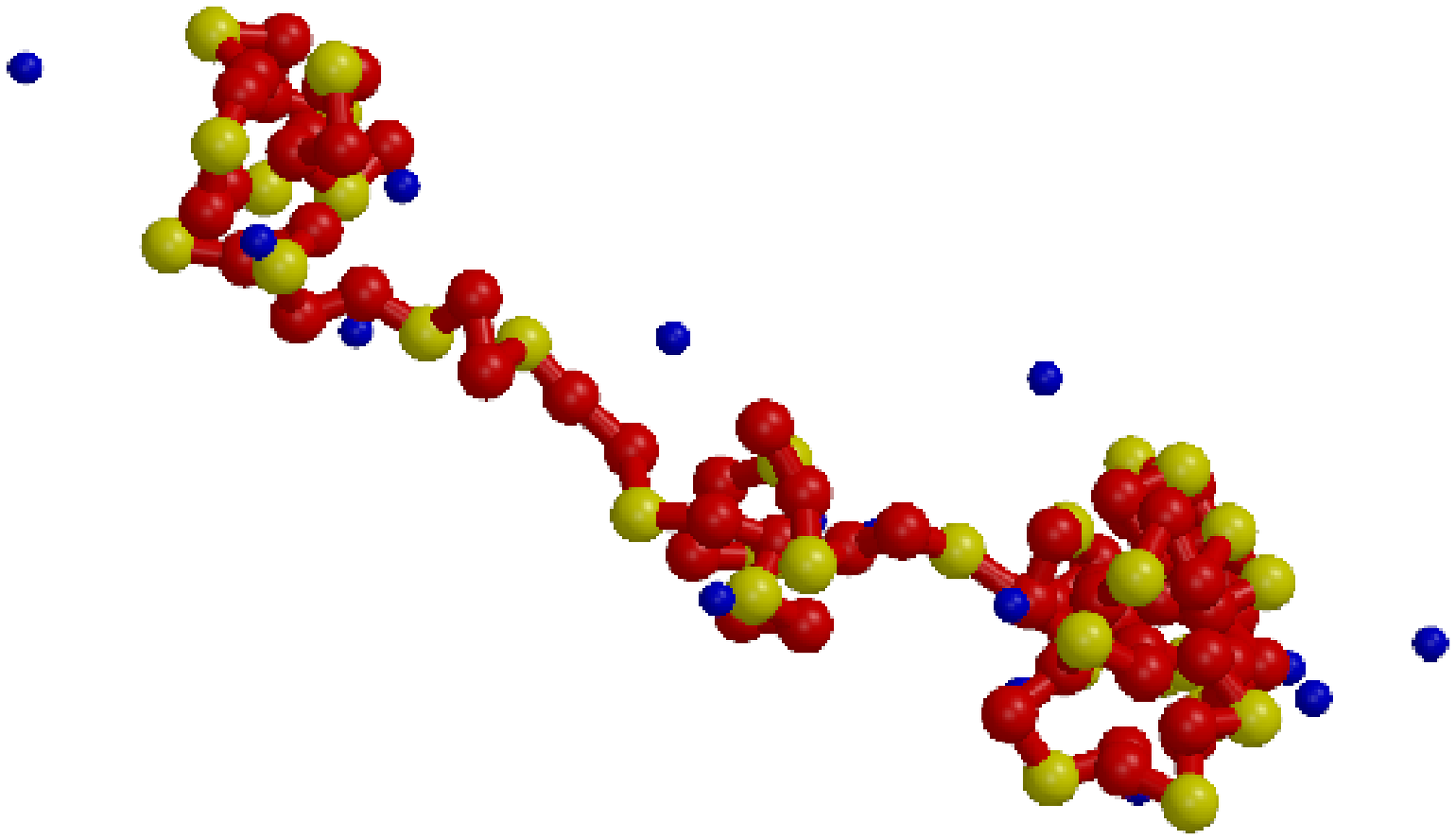,width=7.5 cm}
    \caption{Typical polyelectrolyte conformation for the densities $\rho =
     2\cdot10^{-4} \sigma^{-3}$ (left) and $\rho =2\cdot10^{-5}\sigma^{-3}$ 
     (right). Only counterions within 10 $\sigma$ of the chain are considered.} 
    \label{micka_fig_10-45}
  \end{center}
\end{figure}

The polymer density $\rho$ can be used as a very simple parameter to separate 
different conformation regimes. This can already be seen in the
plots of the end-to-end distance $R_e$ and $r= \frac{R_E^2}{R_G^2}$ 
versus $\rho$ in 
Fig.~\ref{micka_fig_re}. At very high densities
the electrostatic interaction is highly screened, so that the hydrophobic
interaction wins, and the chains collapse to dense globules. If one slightly
decreases the density, the chains can even contract further, because there are
no more steric hinderences from the other chains or counterions, and the
screening is smaller.
The collapsed globules, however, have still a net charge, 
and repel each other, so that this phase
resembles a charged stabilized colloid or microgel phase. 
With decreasing density 
the electrostatic interaction will dominate over the hydrophobic one. 
The chains will tend to elongate, 
assuming pearl-necklace conformations, see Fig.~\ref{micka_fig_10-45}, as they
have been predicted for weakly charged polyelectrolytes in Ref. \cite{neckl}. 
The more the chain stretches, the smaller become the locally compact regions.
Reducing the density even further then leads to a strongly elongated 
blob pole, which is however, due to the intra chain entropy and the 
hydrophobicity, still far from an ideal stiff cylinder, and shows 
local structure.
               \section{CONCLUSION}
We summarized the recent efforts which were pursued to learn more about linear
flexible polyelectrolytes in solution. The investigated systems are already
large enough to reproduce experimental measurements such as the osmotic
pressure or the intrachain structure factor. The incorporation of multivalent
counter ions alters the chain structure drastically. Such aspects can clearly
never be captured by a Debye-H\"uckel approach. The integrated 
ion distribution function
P(r) seems to be a good measure of how one can separate the condensed layer of
counterions from the bulk counterions, and shows clearly that the fraction of
condensed ions is density dependent.

With the beginning investigations
of polyelectrolytes in poor solvents one can also compare the simulations more
easily to experiments. An interesting aspect of hydrophobicity is, that
already weak hydrophobicity can alter the chain structure, and that with
strong hydrophobicity the system can produce a physical gel. The behavior
of $R_E$ versus the density looks very similar to the behavior which is found
for hydrophilic systems with trivalent counterions.

Still many things remain to be done. To name just a few, the whole range
of semi-flexible linear chains is still to be investigated. A detailed study
of Manning condensation is also lacking.
To analyze properly any sort of
aggregation phenomena, one needs to simulate much larger systems. This can, up
to now, only be pursued, if one has an efficiently parallized code, which
incorporates a mesh Ewald method or any other algorithm whose CPU-time 
scales linearly with the total number of charges $N_{tq}$.
               \section{ACKNOWLEDGMENTS}
The authors received financial support through the DFG, and a large
computer grant by the HLRZ J\"ulich under contract hkf06.
Most of these results were obtained, at the various stages, in collaboration
with M. Stevens, U. Micka, and M. Deserno, whose contribution we 
gratefully acknowledge.
               
\end{document}